\documentclass[amsmath,amssymb,superscriptaddress,twocolumn,showpacs,prl,
tightenlines]{revtex4}

\usepackage{longtable,dcolumn,epsfig,graphicx}

\begin{document}
 
\title{Nuclear deformation and neutron excess as competing effects for pygmy
       dipole strength}

\author{R.~Massarczyk} 
\affiliation{Helmholtz-Zentrum Dresden-Rossendorf, 01328 Dresden, Germany}
\affiliation{Technische Universit\"at Dresden, 01062 Dresden, Germany}

\author{R.~Schwengner} 
\affiliation{Helmholtz-Zentrum Dresden-Rossendorf, 01328 Dresden, Germany}

\author{F.~D\"{o}nau} 
\affiliation{Helmholtz-Zentrum Dresden-Rossendorf, 01328 Dresden, Germany}

\author{S.~Frauendorf}
\affiliation{University of Notre Dame, Notre Dame, Indiana 46556, USA}

\author{M.~Anders} 
\affiliation{Helmholtz-Zentrum Dresden-Rossendorf, 01328 Dresden, Germany}
\affiliation{Technische Universit\"at Dresden, 01062 Dresden, Germany}

\author{D.~Bemmerer} 
\affiliation{Helmholtz-Zentrum Dresden-Rossendorf, 01328 Dresden, Germany}

\author{R.~Beyer} 
\affiliation{Helmholtz-Zentrum Dresden-Rossendorf, 01328 Dresden, Germany}

\author{C.~Bhatia}
\altaffiliation{Present address: McMaster University, Hamilton, Ontario,
Canada.}
\affiliation{Duke University, Durham, North Carolina 27708, USA}
\affiliation{Triangle Universities Nuclear Laboratory, Durham, 
	   North Carolina 27708, USA}

\author{E.~Birgersson} 
\altaffiliation{Present address: Areva NP GmbH, 91052 Erlangen,
        Germany.} 
\affiliation{Helmholtz-Zentrum Dresden-Rossendorf, 01328 Dresden, Germany}

\author{M.~Butterling} 
\affiliation{Helmholtz-Zentrum Dresden-Rossendorf, 01328 Dresden, Germany}
\affiliation{Martin-Luther Universit\"at Halle-Wittenberg, 06099 Halle,
       Germany}

\author{Z.~Elekes} 
\affiliation{Helmholtz-Zentrum Dresden-Rossendorf, 01328 Dresden, Germany}

\author{A.~Ferrari} 
\affiliation{Helmholtz-Zentrum Dresden-Rossendorf, 01328 Dresden, Germany}

\author{M.\,E.~Gooden}
\affiliation{Triangle Universities Nuclear Laboratory, Durham, 
	   North Carolina 27708, USA}
\affiliation{North Carolina State University, Raleigh, North Carolina 27695,
       USA}

\author{R.~Hannaske} 
\affiliation{Helmholtz-Zentrum Dresden-Rossendorf, 01328 Dresden, Germany}
\affiliation{Technische Universit\"at Dresden, 01062 Dresden, Germany}

\author{A.\,R.~Junghans} 
\affiliation{Helmholtz-Zentrum Dresden-Rossendorf, 01328 Dresden, Germany}

\author{M.~Kempe} 
\affiliation{Helmholtz-Zentrum Dresden-Rossendorf, 01328 Dresden, Germany}
\affiliation{Technische Universit\"at Dresden, 01062 Dresden, Germany}

\author{J.\,H.~Kelley}
\affiliation{Triangle Universities Nuclear Laboratory, Durham, 
	   North Carolina 27708, USA}
\affiliation{North Carolina State University, Raleigh, North Carolina 27695,
       USA}

\author{T.~K\"ogler} 
\affiliation{Helmholtz-Zentrum Dresden-Rossendorf, 01328 Dresden, Germany}
\affiliation{Technische Universit\"at Dresden, 01062 Dresden, Germany}

\author{A.~Matic} 
\altaffiliation{Present address: IBA Particle Therapy, 45157 Essen, Germany.}
\affiliation{Helmholtz-Zentrum Dresden-Rossendorf, 01328 Dresden, Germany}

\author{M.\,L.~Menzel} 
\altaffiliation{Present address: Max Planck Institut f\"ur Extraterrestrische
Physik, 85741 Garching, Germany.}
\affiliation{Helmholtz-Zentrum Dresden-Rossendorf, 01328 Dresden, Germany}

\author{S.~M\"uller} 
\affiliation{Helmholtz-Zentrum Dresden-Rossendorf, 01328 Dresden, Germany}

\author{T. P.~Reinhardt} 
\affiliation{Technische Universit\"at Dresden, 01062 Dresden, Germany}
\affiliation{Helmholtz-Zentrum Dresden-Rossendorf, 01328 Dresden, Germany}

\author{M.~R\"{o}der} 
\affiliation{Helmholtz-Zentrum Dresden-Rossendorf, 01328 Dresden, Germany}
\affiliation{Technische Universit\"at Dresden, 01062 Dresden, Germany}

\author{G.~Rusev}
\altaffiliation{Present address: Los Alamos National Laboratory, Los Alamos,
        New Mexico 87545, USA.}
\affiliation{Duke University, Durham, North Carolina 27708, USA}
\affiliation{Triangle Universities Nuclear Laboratory, Durham, 
	   North Carolina 27708, USA}

\author{K.\,D.~Schilling} 
\affiliation{Helmholtz-Zentrum Dresden-Rossendorf, 01328 Dresden, Germany}
\affiliation{Technische Universit\"at Dresden, 01062 Dresden, Germany}

\author{K.~Schmidt} 
\affiliation{Helmholtz-Zentrum Dresden-Rossendorf, 01328 Dresden, Germany}
\affiliation{Technische Universit\"at Dresden, 01062 Dresden, Germany}

\author{G.~Schramm} 
\affiliation{Helmholtz-Zentrum Dresden-Rossendorf, 01328 Dresden, Germany}
\affiliation{Technische Universit\"at Dresden, 01062 Dresden, Germany}

\author{A.\,P.~Tonchev}
\altaffiliation{Present address: Lawrence Livermore National Laboratory,
Livermore, California 94550, USA.}
\affiliation{Duke University, Durham, North Carolina 27708, USA}
\affiliation{Triangle Universities Nuclear Laboratory, Durham, 
	   North Carolina 27708, USA}

\author{W.~Tornow}
\affiliation{Duke University, Durham, North Carolina 27708, USA}
\affiliation{Triangle Universities Nuclear Laboratory, Durham, 
	   North Carolina 27708, USA}

\author{A.~Wagner} 
\affiliation{Helmholtz-Zentrum Dresden-Rossendorf, 01328 Dresden, Germany}

\date{\today}

\begin{abstract} 
The electromagnetic dipole strength below the neutron-separation energy has
been studied for the xenon isotopes with mass numbers $A$ = 124, 128, 132, and
134 in nuclear resonance fluorescence experiments using the $\gamma$ELBE 
bremsstrahlung facility at Helmholtz-Zentrum Dresden-Rossendorf and the
HI$\gamma$S facility at Triangle Universities Nuclear Laboratory Durham. 
The systematic study gained new information about the influence of the neutron
excess as well as of nuclear deformation on the strength in the region of the
pygmy dipole resonance. The results are compared with those obtained for the
chain of molybdenum isotopes and with predictions of a random-phase
approximation in a deformed basis. It turned out that the effect of nuclear
deformation plays a minor role compared with the one caused by neutron excess.
A global parametrization of the strength in terms of neutron and proton numbers
allowed us to derive a formula capable of predicting the summed $E1$ strengths
in the pygmy region for a wide mass range of nuclides.
\end{abstract}


\maketitle

Photon strength functions (PSF) are important inputs for statistical reaction
codes applied in network calculations in nuclear astrophysics and in
simulations done for nuclear power production and safety. Knowing and 
understanding the behavior of the PSF in the
energy region around and below 
the neutron threshold is essential for these applications. 
For the dominating electric dipole ($E1$) part of the PSF, the RIPL3 compilation
of the IAEA \cite{cap09} offers an overview on various
models, which in essence base on the concept of the damped isovector $E1$
giant dipole resonance (GDR). It is described by one or two Lorentzian
functions with parameters fitted to the characteristic resonance structure
observed in $(\gamma,n)$ reactions. For open shell nuclei, which constitute the
majority,
the nuclear deformation is taken into account. It  
splits the peak of the GDR \cite{boh75,eis75} and, as a consequence, increases 
the dipole strength distribution in the region
below the neutron-separation energy. Along these lines, a new global
description of the PSF was recently presented in Ref.~\cite{jun08}, which
takes triaxial quadrupole deformation into account and which is called triple
Lorentzian model (TLO). 

Experimental and theoretical studies \cite{lan71,vol06,zil02,ton10,tso04}, which
have been recently reviewed by Savran, Aumann, and Zilges \cite{sav13} suggest a
richer structure of the PSF below the neutron-separation energy than accounted
for by the Lorentzian-type models. In this letter we follow the suggestion in
the review \cite{sav13}: 
"Today the term Pygmy Dipole Resonance (PDR) is frequently used for the
low-lying E1 strength and we will follow this notation in this review without
implying with this notation any further interpretation of its structure." The
rational behind this terminology is that the interpretation of the PDR depends
strongly on the theoretical model invoked and present day experiments cannot
distinguish between the models.\\
One important aspect of studying the PDR concerns its isospin
dependence, which is particularly important for simulating the r-process that
drives the element synthesis in the cosmic evolution. 
Experiments including chains of isotopes with changing ratio of 
neutron-to-proton numbers $N/Z$ address this question. Detailed studies of
isotones with the closed
neutron shells $N =$ 50 \cite{sch13} and $N =$ 82 \cite{sav08}, respectively,
were performed using $(\gamma,\gamma')$ reactions with bremsstrahlung at the
facilities in Dresden and Darmstadt. These studies of spherical nuclides 
 suggested an increase of the pygmy strength with neutron excess. To 
investigate the effect of nuclear deformation on pygmy strength we studied
the even-even Mo isotopes. We found an
increase with $N/Z$ \cite{rus09}, which we attributed to 
the increasing deformation. However, because in the Mo chain 
the deformation increases along with the neutron excess, one cannot disentangle
the two mechanisms.\\

In this Letter, we present the results of a study of the chain of stable
even-mass xenon isotopes. In this series of nuclides, the deformation is
largest for the lightest isotope and decreases with growing neutron number.
We analyze the balance of the effect of nuclear deformation and neutron excess
on the strength in the PDR region for the first time and suggest a
phenomenological model to describe global trends. The data are compared with
calculations in the framework of the deformed Quasiparticle Random Phase (QRPA)
approach. Data of the dipole strength in xenon isotopes only exist for the
energy region
below 4 MeV \cite{gar06} so far, but neither for the PDR nor for the GDR region,
which are also interesting as particular Xe isotopes (e.g. large neutron
capture cross section on $^{135}$Xe with reasonably short half life) are
important as reactor poison for nuclear technology and for the description
of nuclear processes in the solar system \cite{kun98,muk12}.\\

The experiments were performed at the bremsstrahlung facility $\gamma$ELBE of
the Helmholtz-Zentrum Dresden-Rossendorf \cite{sch05}. Electrons with an energy
$E_{\rm{kin}} = $ 12 MeV hit a 7-$\mu$m thick niobium foil and produce
bremsstrahlung with maximum energies exceeding the respective 
neutron-separation energies. The $\gamma$-ray detection setup consisted of four
high-purity germanium detectors surrounded by scintillation detectors acting as
escape-suppression shields. The targets consisted of xenon gas of about 2 g,
highly-enriched (99.9\%) in the respective isotope, at a pressure of about 80
bar in sphere-shaped steel containers \cite{rup09} with an inner diameter of
2 cm. They were combined with disks of 0.2 g of $^{11}$B used for photon flux
calibration.

\begin{figure}[t]
\centering
\includegraphics[width=0.7\columnwidth,angle=270,bb=103 44 574
660,keepaspectratio=true]{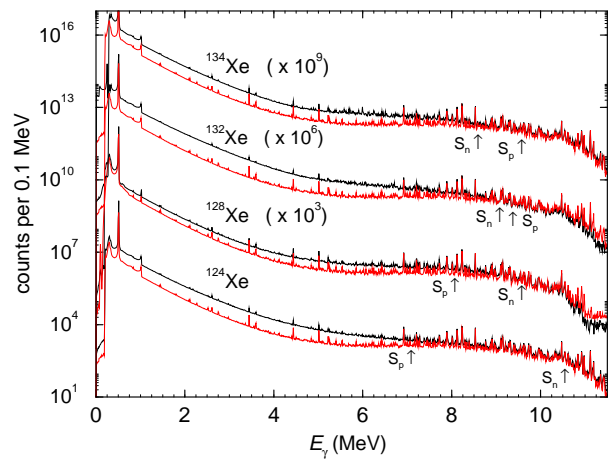}
\caption{(Color online) Measured spectra for filled (black) and empty steel
containers (red). The spectra of $^{134}$Xe, $^{132}$Xe and $^{128}$Xe are
scaled up for a better view. The neutron ($S_n$) and proton ($S_p$) separation
energies are marked with arrows.}
\label{Fig1}
\end{figure}

The first important step in the data analysis was the subtraction of the
background originating from reactions of photons with the steel container. A
comparison of spectra measured for empty and xenon-filled containers is shown
in Fig.\,\ref{Fig1}. The spectra of empty and filled containers were normalized
by adjusting the peak areas of intense transitions in $^{56}$Fe. The
intensities depend only on the integrated photon flux and are therefore
independent of beam fluctuations. The normalization factor derived in this way
fits the one deduced from transitions in $^{11}$B with well known
cross sections \cite{ajz90,rus09b}.

The analysis included the following steps: 
(i) Unfolding of detector response and efficiency using GEANT4 \cite{ago03}, 
(ii) Simulation of the non-nuclear beam-induced background with GEANT4, 
(iii) Correction for inelastic scattering and estimate of branching ratios
using the code $\gamma$DEX for the simulation of statistical $\gamma$-ray
cascades \cite{sch12,mas13}.
A step-by-step review of the data analysis and uncertainty estimate can be
found in Ref.~\cite{mas12}. The total uncertainties are larger compared to
previous measurements with solid targets due to the fact that the
spectrum of the empty container has to be subtracted.

\begin{figure}[t]
\centering
\includegraphics[width=0.7\columnwidth,angle=270, bb=103 31 573 672,
keepaspectratio=true]{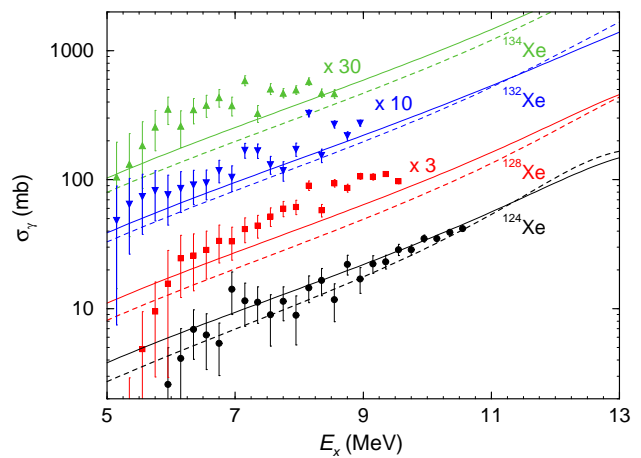}
\caption{(Color online) Resulting photo-absorption cross sections for
$^{124}$Xe (black circles), $^{128}$Xe (red squares) $^{132}$Xe (blue triangles
down), and $^{134}$Xe (green triangles up). For comparison predictions of RIPL3
(solid lines) and the TLO (dashed lines) are shown in the
corresponding colors. The data for $^{128}$Xe, $^{132}$Xe, and $^{134}$Xe are
multiplied by the factors 3, 10 and 30, respectively. }
\label{Fig2}
\end{figure}

The photo-absorption cross sections deduced from the experiments at $\gamma$ELBE
are shown in Fig.\,\ref{Fig2}. These measured cross sections show an enhancement
over the predictions of a two Lorentzian curves, as provided in the RIPL3
database \cite{cap09} and the triple Lorentzian model \cite{jun08} in the
PDR region.

In addition to these experiments we studied the isotopes $^{128}$Xe and
$^{134}$Xe in experiments at the High-Intensity $\gamma$-ray Source
(HI$\gamma$S) \cite{wel09} operated by the Triangle Universities Nuclear
Laboratory (TUNL) in Durham, North Carolina. The main aim of these experiments
using highly polarized quasi-monoenergetic $\gamma$ beams was to identify the
contributions of $E1$ and magnetic dipole $M1$ strength to the total
absorption cross section. The summed $M1$ strength in the energy region between
6 and 8 MeV was less than 10\% of the $E1$ strength in the same energy region.
The summed magnetic dipole strength between 6 and 8 MeV is $B(M1)$ = 0.14(3)
$e^2$ fm$^2$ for $^{128}$Xe and $B(M1)$ = 0.07(2) $e^2$ fm$^2$ for $^{134}$Xe.\\
To study a possible effect of nuclear deformation on the dipole strength, we
performed calculations in the framework of a quasiparticle-random-phase
approximation (QRPA). These calculations take into account quadrupole and
triaxial deformation and are therefore suitable for the present task. The QRPA
is based on a Wood-Saxon mean field and an isovector dipole-dipole interaction
\cite{doe11}. For the calculation of the low-energy part of the PSF the
suppression of the spurious center-of-mass motion was performed
as described in Ref.~\cite{doe05}.

\begin{figure}[h]
\centering
\includegraphics[width=0.7\columnwidth,angle=270,bb=90 50 576 677,
keepaspectratio=true]{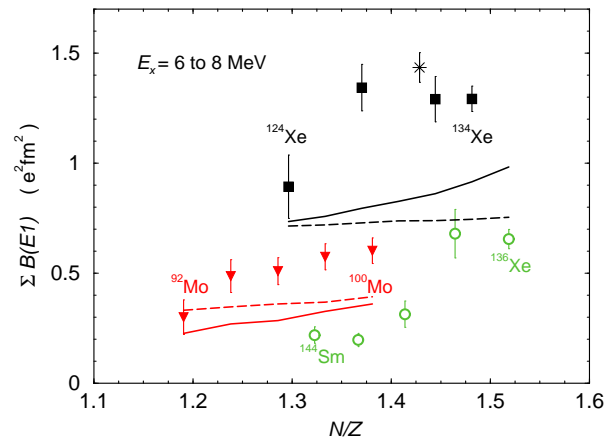}
\caption{(Color online) Summed $B(E1)\uparrow$ values versus neutron-to-proton
number. The $\gamma$ELBE data for the four xenon isotopes (black squares) and
for $^{136}$Ba (black asterisk) are shown in comparison with the $\gamma$ELBE
data for the molybdenum isotopes \cite{rus09} (red triangles) and with results
for $^{136}$Xe and further $N=$ 82 isotones from other experiments \cite{sav08}
(open green circles). Further, the results of the QRPA calculations for all
stable even-mass xenon (black solid line) and molybdenum isotopes (red solid
line) as well as the TLO predictions (dashed lines) are shown.}
\label{Fig3}
\end{figure}

We transformed the experimental absorption cross-sections $\sigma_\gamma$ to
reduced transition strengths $B(E1)$ using the relation \cite{boh75}
\begin{equation} 
4.03 E_x B(E1) \uparrow = \int_{\varDelta E} \sigma(E_x) dE
\approx \sigma_{\gamma}(E_x) \varDelta E 
\label{BE1}
\end{equation}
with $B(E1)$ in $e^2$ fm$^2$, the excitation energy $E_x$ in MeV, and the
absorption cross section $\sigma_\gamma$ in mb in an energy bin $\varDelta E
=$0.2\,MeV.
In the following we will discuss the summed $B(E1)$ strength in the region of
pygmy strength \cite{tso04}, calculated from the interval 6 to 8 MeV, (both for
theory and experiment).\\
Fig.\,\ref{Fig3} shows the summed strengths derived in this way for the Xe
isotopes together with the Mo isotopes and the $N=82$ isotones. The general
experimental trend is an increase of strength with the ratio $N/Z$.
In the case of the Xe isotopes there is almost no change between the heavier
isotopes but a remarkable decrease toward the lightest isotope. The increase
with $N/Z$ was found in our earlier study of Mo isotopes \cite{rus09} already. 

In addition, we show the $B(E1)$ strengths for $N$ = 82 isotones 
calculated from the data presented in
Ref.\,\cite{sav08}. The values show a similar increase with $N/Z$,
but are considerably smaller than our values for nuclides in the mass-130
region. The reason for this difference is that the values of Ref.\,\cite{sav08}
only include strength found in resolved peaks. In contrast, the strengths
resulting from our analyses include strength in the quasi-continuum of states
as well, which amounts to about 60\% of the total strength
\cite{mas12,mak10,rus08,sch08}. The smooth connection of the absorption cross
section derived in this way with the one known from $(\gamma,n)$ experiments at
the neutron threshold proves that it is important to include the strength from
the quasi-continuum. Assuming the data of Ref. \cite{sav08} 
underestimate the total strength by the same factor, the general trend of
increase is also apparent.\\
Fig. \ref{Fig3} compares the experimental strength with the calculated QRPA and
TLO values. The details of the QRPA calculations are described in
Ref.~\cite{doe11, zha09}, where the deformation parameters $\beta_2$, which
result from the micro-macro mean field calculations, are quoted. These values
are consistent with the experimental ones given in Ref.~\cite{ram01} and used
also as input deformation of the TLO model.
The QRPA calculations reproduce the gross dependence on $N/Z$. However, the
scale of pygmy strength is underestimated, which seems to be a general problem 
with our version QPRA that already came up in our earlier studies 
\cite{mak10,mas12}.
At this point, we can only speculate about 
the reason. One possibility is a certain degree of collectivity 
in the pygmy region (the celebrated vibration of the neutron skin
against the core with equal neutron and proton numbers \cite{tso04}),
which is not accounted for by our simple dipole-dipole interaction.
Another possibility are complex excitations beyond the QRPA, as for example
fragmented  two-phonon quadrupole-octupole states. 
The TLO values, which depend very weakly on $N/Z$, are also too low.
The similar scale of QRPA and TLO may not be by accident. Both models
incorporate the isovector dipole vibrations,
obey the TRK $E1$ sum rule \cite{tho25,rei25,kuh25}, and introduce a spreading
width that ensures that the width of the GDR peak is reproduced (see
\cite{jun08,zha09}). 

\begin{figure}[t]
\centering
\includegraphics[width=0.7\columnwidth,angle=270,bb=92 50 576 677,
keepaspectratio=true]{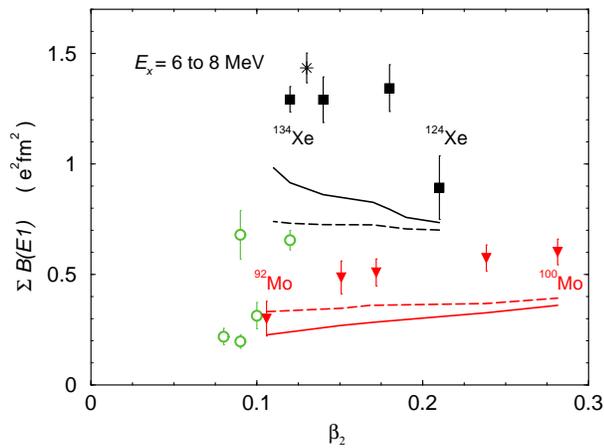}
\caption{(Color online) Summed $B(E1)\uparrow$ values for different isotopes
in the $N=$ 50 and $N=$ 82 region over the nuclear quadrupole deformation
parameter $\beta_2$. The definition of the symbols corresponds to the one in 
Fig.\,\ref{Fig3}.}
\label{Fig4}
\end{figure}
In order to expose the effects of the nuclear quadrupole deformation the 
pygmy strengths are replotted in Fig.\,\ref{Fig4} versus the deformation
parameter $\beta_2$. The Mo chain shows a smooth increase of
pygmy strength with deformation. In this case the growing deformation combines
constructively with the increasing neutron excess. The QRPA well accounts for 
 the increase, whereas the TLO underestimates it. In the case of the Xe
isotopes the combination is destructive.
The values reach from $\beta_2 \approx 0.21$ for $^{124}$Xe with the smallest
$N/Z$ to $\beta_2 \approx 0.1$ for the heavy isotopes. 
The Xe isotopes show a decrease of the pygmy strength with deformation: 
 almost constant values for the nuclides with little deformation and a smaller
value for the deformed $^{124}$Xe, which is the most remarkable finding of this
Letter. The QRPA calculations qualitatively reproduce the decrease with
increasing deformation. The TLO values stay rather constant with a tiny decrease
toward larger deformation. The latter is the result of the competition between
an increase caused by deformation and a decrease caused by the $NZ/A$ dependence
of the TRK on the neutron excess. The QRPA strengths are much more sensitive to
deformation than the TLO ones, which turns out to be important for reproducing
the experimental trends. This indicates that the shell structure, which is taken
into account in QRPA but not in TLO, strongly influences the deformation
dependence of the $E1$ strength.\\
As a result of this analysis, we conclude that the deformation has only a minor
effect on the summed $E1$ strength in the PDR region and that it is sensitive to
the local shell structure. This suggests that the
neutron excess is the key element for the global $N-Z$ dependence of the
strength below the neutron separation threshold. So far,
it has been studied only for spherical nuclei, in experiments as
well as in RQTBA \cite{mas12}, QRPA and QPM\cite{sch13} calculations. To
quantify the global
trends, we  parametrize the integrated $E1$ strength in terms of $N$ and $Z$
neglecting the dependence on the deformation. Fig.\,\ref{Fig3} suggests an
approximately linear dependence on the ratio $N/Z$ 
\begin{equation}
\sum B(E1) \propto \,\left( \frac{N}{Z}-1\right). 
\label{BE1equation}
\end{equation}
The ratios of the pygmy strength between the different isotopes 
are well described by the $N$ - $Z$ dependence of the TRK sum
rule, $NZ/A$. Combining these observation results in the expression
 \begin{equation}
\sum_{6-8 \text{MeV}} B(E1) \approx r \frac{NZ}{A}
\left(\frac{N}{Z}-1\right) 
\label{newBE1equation}
\end{equation}

for the summed strength in $e^2$fm$^2$ in the interval from 6 to 8 MeV.
 To check the quality of the approximation and fix the parameter $r$, we
calculated the ratios $r_{E1}$
of the experimental and parametrized pygmy strengths
\begin{equation}
r_{E1} = \sum B(E1) \left[\left(\frac{N}{Z} - 1\right) 
\frac{N Z}{A}\right]^{-1}.
\end{equation}
These ratios stay rather constant for the considered Mo, Xe and
Ba isotopes around an average value of $r_{E1}$ = 0.080(5),
which determines the free parameter in Eq. (\ref{newBE1equation}) to
$r\approx0.08$ in Eq.
(\ref{newBE1equation}).

\begin{figure}[h]
 \includegraphics[width=0.7\columnwidth,angle=270,bb=61 60 545 679
,keepaspectratio=true]{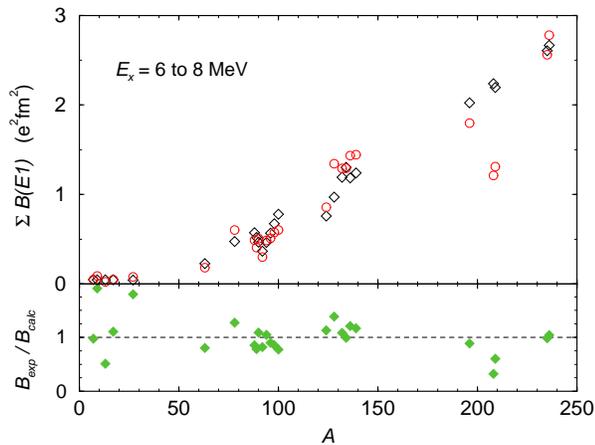} 
 \caption{(Color online) Comparison of experimental data (red circles) and the
prediction by Eq.\,(\ref{newBE1equation}) (black diamonds). The bottom subfigure
shows the divergence of the two.}
 \label{Fig5}
\end{figure}

We tested Eq.\,(\ref{newBE1equation}) for nuclides with masses up to 250 using
experimental data available in the EXFOR database \cite{exfor}. The comparison
of the predictions of Eq.\,(\ref{newBE1equation}) with experimental $B(E1)$
values deduced from the cross sections given in EXFOR according to 
Eq.\,(\ref{BE1}) is shown in Fig.\,\ref{Fig5}. We find a generally good
agreement in the whole considered mass region including light nuclides as well
as very heavy nuclides such as uranium and thorium isotopes which are well
deformed. The general agreement supports our finding that deformation has a 
minor influence on the low-lying dipole strength on the global scale. 

As a caveat, we notice that the summed $E1$ strength of double magic nuclei,
such as $^{208}$Pb, cannot be described with Eq.\,(\ref{newBE1equation}),
probably because of their exceptional small fragmentation of the low-lying
strength (see also Ref.~\cite{sch10}). Further, the formula fails for nuclides
with $N = Z$. For practical purposes, this is of minor importance because the
valley of stability does not follow the $N=Z$ line and applications in nuclear
technology as well as in nuclear astrophysics concern mainly neutron-rich
nuclei. 

Summarizing, we analyzed the dipole strength in the PDR region in the chain of
xenon isotopes. In contrast to our earlier study of molybdenum isotopes we could
disentangle the effects of neutron excess and nuclear deformation on the dipole
strength and study their competition for the first time. We found the neutron
excess to have the dominating effect on the strength below the neutron
separation energy. We constructed a simple parametrization of the summed PDR
strength that describes its global $N$ and $Z$ dependence in a wide mass range
of nuclides.\\

We thank A. Hartmann for technical support and the crews of the ELBE
accelerator and of the HI$\gamma$S facility for their collaboration during the
experiments. 
This work was supported by the German Research Foundation (DFG), project no.
SCHW883/1-1 and partly by the EURATOM FP7 Project ERINDA (FP7-269499).
Partial support was also granted by the U.S. Department of Energy, Office of
Nuclear Physics, under Grants No. DE-FG02-95ER4093, No. DE-FG02-97ER41033
and No. DE-FG02-97ER41042.

\end{document}